**A Phase-Field Study on the Effects of Nanoparticle Distributions on Solidification and Grain Growth**

**Bryan Kinzer[1*], Rohini Bala Chandran[1*]**

Department of Mechanical Engineering, University of Michigan, Ann Arbor, MI 48109, USA[1]

**\*Corresponding Authors, brkinzer@umich.edu, rbchan@umich.edu**

Abstract

Nanoparticle reinforced alloys offer the potential of high strength, high temperature alloys. While promising, during rapid solidification processes, alloys suffer from nanoparticle clustering, which can discount any strength benefit. An open-source phase-field model is developed using PRISMS-PF to explore the impact of nanoparticles and clustering on alloy solidification. Heterogenous nucleation and grain boundary pinning are explicitly included, and a wide range of nanoparticle area fractions (0.01–0.1) and nucleation rates ($10^6$–$10^{12}$nuclei/m$^2$) are modeled. At low area fractions (< 0.05), particle clustering increases grain size between 15-45% compared to a random distribution. Our quantitative analyses inform a modified Zener grain size relationship that not only depends on nanoparticle size and area fraction, but also on the nucleation rate. Grain size first drastically decreases before plateauing at higher nucleation rates. Our simulations reveal a strong preference of nanoparticles pinning grain boundaries. Pinning fraction increases rapidly with nucleation rate before saturating between 0.85–0.90. Across the range of area fractions and nucleation rates considered, the random and clustered grain sizes each collapse to a simple analytical expression that depends only on nanoparticle radius and pinning fraction. Comparisons against experimental data reveal the expressions deduced from our analyses fit reported grain sizes better than classic Zener analysis. A simple model of strength and cost tradeoffs indicates nanoparticles can be a cost-effective way to improve alloy strength.

## 1. Introduction

Alloys containing a dispersion of nanoparticles offer the potential to enable next generation high temperature, high strength alloys [1–5]. Embedding ~10–100 nm ceramic particles, such as TiC, $TiB_2$, $Y_2O_3$, and $Al_2O_3$, has been demonstrated to result in a near doubling of yield stress and improved creep performance from room temperature to elevated temperatures (≥ 800 °C) as compared to standard particle-free alloys [6–9]. Metal matrix nanocomposite alloys, such as Ni-based MA6000 and ferrous Kanthal APM, have been manufactured through conventional powder metallurgy involving mechanical alloying with subsequent hot rolling and extrusion [5,10,11]. Laser additive manufacturing (AM) is an emerging manufacturing technique to fabricate complex geometries with metal alloys [1,12–14]. While promising, development of nanocomposites is hindered by a tendency of nanoparticles to cluster/agglomerate during techniques, such as selective laser melting (SLM), due to rapid solidification of the melt pool [13,15,16]. Clustering causes inhomogeneities in the microstructure and weakens the alloy [16]. The focus of our study is to use phase-field modelling to explore the impact of nanoparticle distributions on the microstructure evolution of SLM alloys both during and after solidification.

Addition of nanoparticles improves alloy mechanical properties by reducing average grain size and preventing grain growth. However, the strength behavior of SLM alloys can be limited if longer, larger columnar grains form during solidification, which leads to inferior, anisotropic mechanical properties as compared to equiaxed grains [17,18]. It has been demonstrated computationally and experimentally that nanoparticles serve as grain nucleation sites. The nanoparticles can arrest columnar grain growth resulting in a finer, more equiaxed microstructure capable of achieving grain sizes on the order of a few microns to a few hundred nanometers [6,19–21]. In this grain size range further size decreases can have a positive impact to improve yield strength by increasing the density of grain boundaries that serve as obstacles for dislocations [9]. In the absence of nanoparticles, at elevated temperatures the resistance to dislocations decreases resulting in grain boundary sliding of the metal matrix. This reduces both yield and creep strength as the metal slowly deforms over time [22]. Thus, even with initially small grain sizes, there is a thermodynamic driving force that causes grain boundaries to grow to reduce free energy, which is further exacerbated at higher temperatures.

Even with major benefits associated with improved strength properties from nanoparticles, the laser AM process for manufacturing nanoparticle reinforced metal alloys is limited by clustering of particles. D. Gou et al. have reported that severe clustering of nanoparticles can reduce the elastic modulus of TiC reinforced Inconel 718 by a factor of two, as compared to a uniform dispersion of particles [13]. Clustering has also been shown to lead to detrimental columnar grains with anisotropic mechanical properties [17,18]. While it is relatively easy to achieve a uniform dispersion of microparticles, nanoparticles have a higher tendency to cluster [20,23]. Clustering of nanoparticles is a complex phenomenon driven by combined effects of van der Waals forces, Marangoni convection, and solidification of the melt [13,20,21,24,25]. During SLM the melt pool at the center of the laser has a higher temperature than the laser edge. This can lead to a larger surface tension value at the edge of the laser compared to the center, which draws nanoparticles from the center to the edge, and thereby counteracts the attractive van der Waals forces [13]. If the surface tension forces exceed the van der Waals forces, it can drive particles towards the liquid-solid metal interface, resulting in clustering of nanoparticles at grain boundaries [20]. There is a need to quantify the impacts of spatial distributions of nanoparticles, including clustering effects, on solidification, microstructure evolution, and alloy properties.

Phase-field modelling is a powerful tool to understand complex microstructure evolution and has been deployed to study grain growth in the presence of nanoparticles exerting Zener pinning of grain boundaries [26–28]. Models have been developed that allow nanoparticles to serve as sites for heterogeneous nucleation [29,30], which demonstrates that the nanoparticles facilitate a columnar-to-equiaxed grain transition during SLM. This is due to grains nucleating from nanoparticles in the bulk of the melt arresting any potential columnar grains during solidification [30]. However, most studies preclude the effects of particle clustering on microstructure evolution. Even if included, these analyses are limited to very small volume/area fractions (<0.001) and/or ignore particle pinning [15,25,31]. Therefore, there is still room for improvement of phase-field models to account for particle clustering to aid in the development of new laser processable metal nanocomposites.

The primary objective of this study is to explore the impact of nanoparticle dispersions on microstructure evolution using phase-field simulations. Nanoparticles are modeled to serve as nucleation sites and exert grain boundary pinning forces. Simulations are performed on a two-dimensional modeling domain to explore the effects of (a) particle distributions (random and clustered), (b) nanoparticle concentration quantified by area fractions (0.01, 0.02, 0.03, 0.05, and 0.10), and (c) nucleation rates on the transient evolution of average grain size and size distributions. Notable advancements in our work compared to prior studies are: (1) a systematic quantification of particle clustering effects on alloy solidification and steady-state grain size; (2) quantification of the dependence of nucleation rate on the steady-state microstructure; (3)

identification of a saturation nucleation rate regime beyond which grain growth levels off; and (4) interpreting results to evaluate trade-offs in strength gains and materials cost as a function of nanoparticle concentration. Our results illustrate that for a wide variety of area fractions, nucleation rates, and particle distributions, the grain size is most strongly dictated by the fraction of nanoparticles on grain boundaries.

## 2. Methods

Two-dimensional (2D) phase-field simulations are conducted using the open-source software PRISMS-PF, which is a parallelized finite element code to determine microstructure evolution [32]. In phase-field simulations, governing equations are formulated to represent thermodynamic driving forces for grain growth, for interactions between grains, and for kinetic parameters that dictate the rate of microstructure evolution. Compared to other modeling approaches, such as cellular automata that explicitly track interfaces [33], phase-field simulations treat interfaces to be diffuse with the order parameters transitioning smoothly across the width of the interface [32]. This feature leads to implicit tracking of complex grain morphologies and is therefore valuable for our problem.

### 2.1 Modeling Domain

Fig. 1a shows the 2D, 5 μm × 5 μm, modeling domain and boundary conditions for the phase-field simulations. The simulations represent the solidification of a single-layer, single-track SLM process, so a substrate is included. This domain size enables tracking several hundred nanoparticles and up to several hundred grains to yield statistically meaningful results while remaining computationally tractable. Typically, reinforcing nanoparticle radii range from 5–

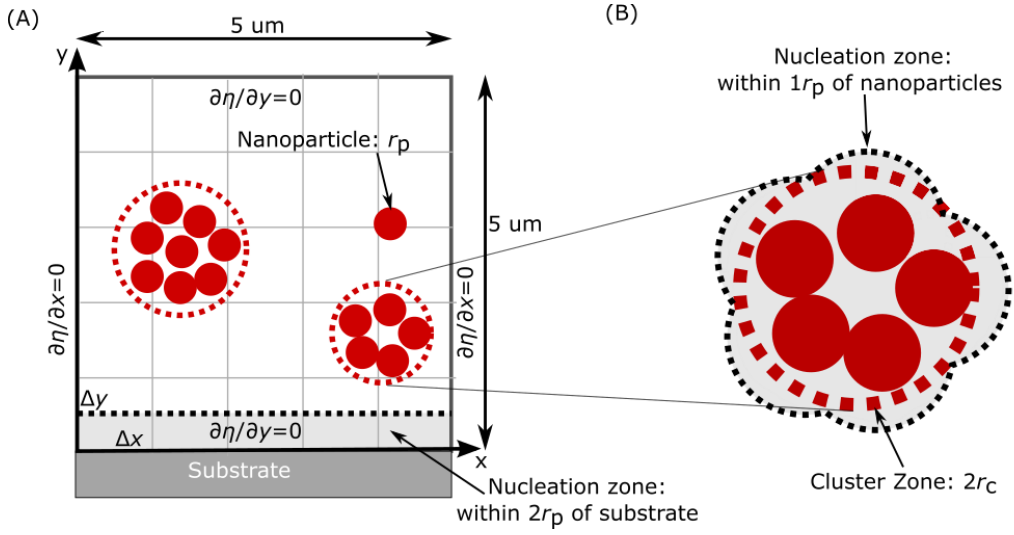

**Fig. 1.** Computational domain. (a) Schematic of the computational domain with a mesh size of 40 nm × 40 nm and with a sample clustered distribution of monodisperse nanoparticles with radius of $r_p$ = 25 nm. Neumann boundary conditions are applied for the order parameter, η, at all boundaries. A constant temperature gradient of $\partial T / \partial y$ = $10^5$ K/m is applied in the positive *y*-direction. (b) Aggregations of individual nanoparticles form clusters based on size distribution data with the cluster zone denoted by the dotted red circle with radius, $r_c$. The shaded region bounded by the dotted black curves within 1 $r_p$ of each nanoparticle and within 2 $r_p$ of the substrate wall represents heterogeneous nucleation zones.

50 nm to have a discernable impact on alloy strength. Therefore, we use a baseline particle radius, $r_p$, of 25 nm for all simulations [13]. Depending on the distribution, nanoparticles are placed into discrete cluster zones (Fig. 1a,b). The inclusion of particle dispersions requires two added functionalities in the standard grain growth model implemented in PRISMS-PF [34]. First, we include the capability of nanoparticles and the substrate to serve as heterogenous nucleation sites during solidification, so that new grains can preferentially solidify around the nanoparticles and the substrate [35,36]. Nucleation can occur within 1 $r_p$ of a nanoparticle (Fig. 1b) or within 2 $r_p$ of the substrate (Fig. 1a,1b). Second, we include grain boundary pinning, as nanoparticles pin the grain boundaries limiting grain growth/coarsening, which reduces the final, steady state grain size [27,30].

## 2.2 Nanoparticle Distributions in the Metal Matrix

We model the effects of two distinct spatial distributions for the nanoparticles: (a) random and (b) clustered. A control case is also modeled without any nanoparticles present in the microstructure [37]. Five area fractions, which is the ratio of the total area occupied by the particles (area per particle is $\pi r_p^2$) to the domain area (5 μm × 5 μm), spanning an order-of-magnitude variation, $\phi_{np}$ = 0.01, 0.02, 0.03, 0.05, and 0.1 are modeled. By invoking the principle of Delesse it can be assumed area and volume fractions are equivalent when the nanoparticle concentration is consistently scaled from 2D to 3D [38]. This range includes typical area fractions used in practical applications of nanoparticle alloys [1,13,37].

For the random distribution, a 25 nm radius particle is placed at each randomly sampled site from a uniform distribution of spatial coordinates within the modeling domain. Particle overlap is prohibited, and new particle centers are generated until the area fraction is reached. To model realistic clusters of nanoparticles in a metal matrix, we applied experimentally measured cluster size distributions by D. Gu and coauthors, where Ni-based Inconel 718 nanocomposite with 25 nm radius TiC particles was manufactured via selective laser melting [13]. To generate the clustered distributions (or agglomerates) of particles in our modeling domain (Fig. 1a), coordinates for the center of the cluster are randomly seeded. Then, the experimentally measured cluster size distribution is used to generate the frequency of cluster zones in the domain (Fig. 1b). Individual 25 nm particles are randomly placed within this zone until no more particles can be placed without causing particle overlap. After one cluster is placed, a new cluster site coordinate is randomly selected, and the cluster size distribution informs the radius of the new cluster zone and particles fill the new zone. This process is repeated until the target area fraction is attained. For the clustered case, the cluster size distribution is obtained from the laser processing settings that resulted in an approximately 275 nm average cluster zone radius, which still yielded structurally stable coupons [13]. The modeled size distribution for the clustered case is presented in Fig. SI.1.

## 2.3 Governing Equations for Phase-field Model

Each grain is assigned an order parameter, $\eta_i$, representing a grain orientation with $i$ ranging from $N$ = 1, 2, …, 16. Grains are represented by discrete areas where $\eta_i$ = 1 within the $i^{th}$ grain orientation and each order parameter smoothly transitions from a value of 1 to 0 across the grain boundary interface. $\varphi$ is an additional variable that represents the nanoparticles equaling 0 inside the nanoparticle and 1 everywhere else and is responsible for the pinning effect.

Microstructure evolution, including grain size and orientation, is dictated by the configuration that minimizes the free energy of the system, $F_{sys}$ (Eq. (1))

$$F_{sys} = \int_V f_0 \, dV, \tag{1}$$

which is applied in every $dV$, differential element volume of the domain. $f_0$ is the free energy density functional consisting of bulk and gradient terms in Eq. (2),

$$f_0 = f_{bulk} + f_{grad}. \tag{2}$$

Eq. (2) arises from assuming the free energy depends both on the local free energy, $f_{bulk}$, as well as the local gradient, $f_{grad}$. $f_{bulk}$ is dictated by the free energies of the grains, whereas $f_{grad}$ penalizes sharp gradients and controls nanoparticle interactions with grain boundaries. Both these free energies act together to penalize the formation of interfaces, which drives grain coarsening where smaller grains are consumed by larger grains to reduce the total interfacial area. $f_{\text{bulk}}$, given by Eq. (3),

$$f_{\text{bulk}} = m_{\text{g}} \left( \sum_{i=1}^{N} \left( -\frac{1}{2}\eta_i^2 + \frac{1}{4}\eta_i^4 \right) \right) + m_{\text{g}} \left( \gamma \sum_{i=1}^{N} \sum_{j\neq i}^{N} (\eta_i^2 \eta_j^2) \right),$$  (3)

is formulated with two terms—a double-well free energy functional and a grain interaction term. The first term has global minima at $\eta_i = \pm 1$ and a maximum at $\eta_i = 0$ [39]. Since the liquid phase corresponds to $\eta_i = 0$, the driving force for converting the initial liquid phase into solid grain growth is to reduce this bulk free energy. The second term is a product of two order parameters, $\eta_i$ and $\eta_j$, and grain interaction coefficient, $\gamma$. Across a grain boundary both $\eta_i$ and $\eta_j$ may be non-zero, so this second term penalizes the physically unrealistic outcome of two grains coexisting in the same location. Within a grain, order parameters remain constant since $\eta_i = \pm 1$ and all other $\eta_j = 0$ meaning there is no free energy driving force. $\gamma$ is set to 1.5, because it ensures that the order parameters transform at a constant, symmetric rate. With this symmetry, the rate of generation of $\eta_i$ is equal in magnitude to the rate of consumption of $\eta_j$, i.e., $\partial \eta_j / \partial \eta_i = -1$ [39]. Both terms in Eq. (3) are scaled by $m_{\text{g}}$ (Eq. (4)), which is an interface coefficient related to the width of the grain boundary,

$$m_{\text{g}} = \frac{3}{4} \frac{\sigma_{\text{g}}}{\Delta f_{\text{g}} l_{\text{g}}},$$  (4)

where $l_{\text{g}}$ is the grain boundary width, and $\sigma_{\text{g}}$ is the grain boundary energy, which is assumed to be the same as the solid-liquid interfacial energy; $\Delta f_{\text{g}}$ is the maximum energy barrier of the normalized bulk free energy, $f_{\text{bulk}}/m_{\text{g}}$, in Eq. (3) [37,39]. As $m_{\text{g}}$ increases, the thermodynamic penalty for having an interface increases, resulting in sharper interfaces. This interface coefficient is hence tuned to ensure stability of the simulations resulting in $m_{\text{g}} = 2.1 \times 10^7$ J/m$^3$ for this study.

The second term of Eq. (2) is the gradient term given by Eq. (5)

$$f_{\text{grad}} = \frac{\kappa_{\text{g}}}{2} \varphi \sum_{i=1}^{N} |\nabla \eta_i|^2. \tag{5}$$

$\kappa_{\text{g}}$ is a positive gradient coefficient (Eq. (6)) [39]

$$\kappa_{\text{g}} = a_{\text{k}} \sigma_{\text{g}} l_{\text{g}}, \tag{6}$$

where $a_{\text{k}}$ is the anisotropic coefficient. $\kappa_{\text{g}}$ scales the local gradient, $\nabla \eta$, so a larger value of $\kappa_{\text{g}}$ penalizes sharp interfaces, making the grain boundary interface more diffuse. $a_{\text{k}}$ is set to 0.15 as this favors grain growth in the build direction along the positive y-axis in Fig. 1, to match experimental results and be consistent with previous simulations [37]. Across a typical grain boundary, the order parameter spatial gradient for any grain orientation is non-zero, i.e., $\nabla \eta_i \neq 0$ [40]. The nanoparticle order parameter, $\varphi$, in Eq. (5) differs from other simulations where an additional term is introduced into $f_{\text{bulk}}$ [30,41]. However, the main requirement is just that $\varphi$ should force a discontinuity in the free energy; this energy penalty prevents pinned grain boundaries from detaching. Inside a nanoparticle, $\varphi = 1$, because the nanoparticles are defined with sharp interfaces, and $\varphi = 0$ immediately outside the nanoparticles inducing the required free energy discontinuity (Eq. (7)).

The evolution of the grain microstructure is obtained by computing the spatiotemporal evolution for each order parameter, $\eta_i$. The Ginzburg-Landau equation, Eq. (7), is derived by observing that the time rate of change of the order parameters is proportional to the slope of the spatial free energy, $f_0$, with respect to the order parameter [42]. Taking the partial derivative of $f_0$ with respect to $\eta_i$ using the terms of Eq. (2), and scaling by $L_{\text{g}}$ leads to

$$\frac{\partial \eta_i(\vec{r},t)}{\partial t} = -L_g \frac{\partial f_0}{\partial \eta_i(\vec{r},t)} = -L_g(m_g[-\eta_i + \eta_i^3 + 2\gamma\eta_i \sum_{j\neq i}^N \eta_j^2] - \varphi\kappa_g\nabla^2\eta_i), \tag{7}$$

where $\vec{r}$ is the spatial coordinate, $t$ is time, and $L_g$ is a constant which depends on the rate at which atoms can move across the interface and is introduced to scale the temperature-dependent rate of grain growth [30]. The nanoparticle positions are fixed in size and position, and therefore there is no time evolution in the positions of these particles. $L_g$ is given by Eq. (8),

$$L_g = \frac{1}{a_k}\frac{1}{l_g} D_0 \exp\left(-\frac{Q_g}{RT}\right), \tag{8}$$

where $D_0$ is a pre-exponential constant for grain boundary mobility, and $Q_g$ is the activation energy for grain boundary mobility.

## 2.4 Grain Nucleation and Remapping

PRISMS-PF deploys classic nucleation theory, which can account for both homogeneous and heterogenous nucleation [43]. While homogenous nucleation may occur anywhere within the liquid melt during solidification, heterogeneous nucleation only occurs at preferential sites of solid surfaces that reduce the free energy to form a nucleus, including the imperfections on the substrate wall and on the nanoparticle surfaces (Fig. 1a). The nucleation rates determine frequency of nuclei/grain generation. When a new nucleus is formed it is randomly assigned an order parameter between $N$ = 1–16. While the nucleus radius will vary with the extent of undercooling, we assume all nuclei are generated at a constant radius of 25 nm for simplicity.

The areal nucleation rate equation, $J^*$, is given by Eq. (9) [37,44]:

$$J^* = N_{\text{sites}}\frac{k_B T}{h} \exp\left(-\frac{Q_D}{RT}\right) \exp\left(-\frac{\Delta G^* \zeta_\theta}{k_B T}\right), \tag{9}$$

where $N_{sites}$ is the areal number density for nucleation sites, and $T$ is the temperature as a function of space and time in the computational grid (Fig. 1a), which is dictated by the cooling rate and spatial temperature gradients. Direct measurement of $N_{sites}$ is difficult, and therefore we evaluate grain size as a function of $N_{sites}$ using a baseline value of $1.95 \times 10^9$ nuclei/m². In the remainder of the text, this baseline value is referred to as $N_0 = 1$ and a range of $10^{-3}N_0$–$10^3N_0$, i.e., $1.95 \times 10^6$–$1.95 \times 10^{12}$ nuclei/m² is explored (Table 1). This range was selected as the number of nucleated grains varies from about a few dozen to a few hundred inside the domain, allowing a wide range of microstructures to be studied while remaining statistically valid. Assuming a dimensional scaling law of (atoms/m²)$^{3/2}$ = atoms/m³, the range of $N_{sites}$ considered in our study becomes equivalent to $10^9$–$10^{18}$ atoms/m³ [45]. In comparison, prior 3D studies have covered ranges from $10^{13}$–$10^{15}$ atoms/m³ [18,30]. The first exponential term in Eq. (9) governs thermally activated atomic diffusion across the solid-liquid interface with $Q_D$ being the activation energy for atomic diffusion. The second exponential term signifies the probability that the nucleus formed acquires a critical size. This probability depends on a contact angle function, $\zeta_\theta$, based on the wettability of the nucleus with the nanoparticle/substrate surface and the energy barrier, $\Delta G^*$, to form critically sized nuclei. The contact angle function, which governs the heterogenous nucleation shape factor, is given by Eq. (10),

$$\zeta_\theta = \frac{1}{4}\big(2 + cos(\theta)\big)(1 - cos(\theta))^2. \tag{10}$$

A constant contact angle of $\theta = 4°$, indicative of low surface tension and extremely good wettability is assumed for a simplified analysis to model heterogeneous nucleation in this study; this results in $\zeta_\theta = 4.4 \times 10^{-6}$. For homogenous nucleation locations in the liquid melt, $\theta = 180°$ and $\zeta_\theta = 1$, which results in homogeneous nucleation rates being insignificant (several orders of magnitude smaller) compared to heterogenous nucleation. Therefore, homogenous nucleation events are not included in the model (Fig. 1a).

In classic nucleation theory, the value of $\Delta G^*$ depends on the assumptions of the nuclei shape [46–48]. Even though simulations are 2D, we assume spherical nuclei that project as circles onto the 2D domain. This hybrid approach allows use of experimentally known parameters to calculate $\Delta G^*$ for 3D nucleation while still allowing the model to remain computationally tractable with 2D grain growth kinetics. Following the spherical nuclei assumption, Eq. (11) yields,

$$\Delta G^* = \frac{16}{3} \pi \frac{(\sigma_p)^3}{\left(H_L \frac{\Delta T}{T_m}\right)^2}, \tag{11}$$

where $\sigma_p$ is the solid-liquid interfacial energy, $H_L$ is the latent heat, and $T_m$ is the alloy melting point. $\Delta T$ is the extent of undercooling, $\Delta T = T_m - T$.

Nucleation events are modeled as a stochastic process to model nuclei development in a random manner governed by the local nucleation rate [47]. The nucleation rate in Eq. (9) is used to inform the probability, $P_{nuc}$, of a single nucleus forming inside every cell of the phase-field modeling domain over a finite time. Nucleation events are modeled as a Poisson distribution occurring at a rate, $J^*$. By assuming the likelihood of the formation of more than one nucleus at any instant of time is negligibly small, the nucleation probability for forming one nucleus over a finite time-period of $\Delta t_{nuc}$ becomes

$$P_{nuc} = 1.0 - \exp\left(-J^* \Delta x \Delta y \Delta t_{nuc}\right), \tag{12}$$

where the exponential term represents the probability of no nucleation occurring in any cell [47]. Eq. (12) is applied to every cell in the domain for every nucleation time step, $\Delta t_{nuc}$, and $\Delta x$ and $\Delta y$ are the mesh sizes. In each cell a random number between 0 and 1 is generated, and if the number is less than $P_{nuc}$ a nucleus will attempt to form. However, if there is already a grain, nucleus, or particle at the attempted location the nucleus will not form. Fig. SI.2 shows the temperature dependent nucleation rate curve assuming $N_0 = 1$ (Eq. (9)). When the density of nucleation sites, $N_{sites}$, changes an order of magnitude the nucleation probability also changes an order of magnitude. To reduce the computational cost, nucleation is only attempted every 50 timesteps of the phase-field (order parameter) model, i.e., $\Delta t_{nuc} = 50\Delta t$. This nucleation timestep is still small enough that grain growth between the phase-field timesteps is not substantial to cover nucleation sites that could have generated nuclei. Because of the stochastic nature of nucleation (Eq. (12)), and therefore microstructure evolution, every simulation is run 3 times to quantify mean and standard deviation of model outputs.

## 2.5 Model Parameters

Stainless steel 316L (SS316L) with 25 nm radius $TiB_2$ nanoparticles is chosen as a model material due to its widespread use in additive manufacturing processes and available modeling input parameters from prior implementation in phase-field models [30,37]. Modeling parameters in Eqs. (1–8) have been adopted from Wang et. al. [37], who studied a similar SS316L system, and are listed in Table 1. The values of $D_0$ and grain boundary width, $l_g$, are adjusted to obtain reasonable and stable results [30,37]. In real systems, $l_g$ is typically on the order of 1 nm, but in most phase-field models $l_g$ is assumed to be multiple orders of magnitude larger to allow the spatial resolution to remain computationally tractable [39,49]. Different assumptions about $D_0$ and $l_g$ may change the physical time of the simulations, but the same trends are expected for the simulated microstructures.

Table 1. Model parameters, their baseline values and ranges explored in the phase-field simulations with relevant references

| Category (Equations) | Model parameter, symbol | Value [Units], Reference |
|---|---|---|
| Ginzburg-Landau equation (Eq. (3–8)) | Grain Interaction Coefficient, $\gamma$ | 1.5, [39] |
| | Anisotropy Coefficient, $a_k$ | 0.15, [37] |
| | Grain Boundary Width, $l_g$ | 110 [nm], tunable, [49] |
| | Grain Boundary Pre-exponential Constant, $D_0$ | $1.3 \times 10^{-5}$ [m$^4$/(J·s)], tunable, [37] |
| | Grain Boundary Energy, $\sigma_g$ | 0.385 [J/m$^2$], [37] |
| | Activation Energy for Grain Boundary Mobility, $Q_g$ | 140 [kJ/mol], [37] |
| | Nucleation Free Energy Barrier, $\Delta f_g$ | 0.125, [39] |
| Nucleation (Eq. (9–12)) | Liquid/Solid Interfacial Energy, $\sigma_p$ | 0.385 [J/m$^2$], [37] |
| | Activation Energy for Diffusion, $Q_D$ | 140 [kJ/mol], [37] |
| | Nuclei Contact Angle, $\theta$ | 4°, tunable, [37] |
| | Latent Heat, $H_L$ | $2.1 \times 10^9$ [J/m$^3$], [37] |
| | Melting Point of SS316L, $T_m$ | 1723 [K], [37] |
| | Areal density of nucleation sites, $N_{sites}$ | $1.95 \times 10^6$–$1.95 \times 10^{12}$ [Nuclei/m$^2$], [37] |
| | Dimensionless nucleation sites, $N_0$ | $N_{sites}/ 1.95 \times 10^9$ [Nuclei/m$^2$] |
| Step Sizes | Mesh Size, $\Delta x = \Delta y$ | 40 nm |
| | Model Time Step, $\Delta t$ | 10 ns |
| | Nucleation Time Step, $\Delta t_{nuc}$ | 50 ns |

## 2.6 Simulation Procedure

All simulations are performed in PRISMS-PF (Fig. 1) with a maximum of 16,384 elements, corresponding to a mesh size of $\Delta x = \Delta y = 40$ nm, and 66049 degrees of freedom [32]. PRISMS-PF deploys a matrix-free finite element method solver to obtain order parameter evolution (Eq. (7)). The code used in the study is open-source and available for use on the PRISMS-PF Github page [50]. To reduce computational complexity, the grain remapping algorithm in PRISMS-PF is deployed [34,51]. Rather than having a unique order parameter for every single grain, multiple grains are assigned to the same order parameter. More details are presented in Appendix II. The PRISMS-PF code is parallelized, and simulations are performed on a high-performance computing cluster (Great Lakes, CentOS 7) using 36 cores (2x 3.0 GHz Intel Xeon Gold 6154, 180 GB). The largest nanoparticle area fraction, and nucleation rate modeled takes a maximum of 1 hour. A reference case of the random distribution of nanoparticles with an area fraction of 0.05, and with a nucleation site density of $N_0$ (standard deviation, $s_{dz}$ = 3.2%) is used in discussing impact of parameter selection.

*Boundary and Initial Conditions:* Neumann boundary conditions are modeled for the order parameters on all the boundaries (Fig. 1a). All the order parameters are initialized with values of 0. Nanoparticle distributions generated based on the area fraction and clustering are loaded into fixed positions in the liquid domain to initialize the simulation.

Prior heat transfer simulations for SLM processes indicate transient cooling rates of $10^5$–$10^6$ K/s and spatial temperature gradients of $10^5$–$10^7$ K/m [1,37]. We assumed an intermediate cooling rate of $5 \times 10^5$ K/s, and a spatial temperature gradient of $10^5$ K/m across the build-direction (y-direction in Fig. 1a), which results in a 0.5 K difference across the domain (i.e., y-direction, in the 2D domain modeled). While varying the thermal gradient from 0 K/m, which is an isothermal domain, ($s_{dz}$ = 1.6%) to $10^6$ K/m ($s_{dz}$ = 0.21%) has some impact on when different portions of the domain will nucleate on nanoparticles, the final microstructure still remains isotropic across all situations. While in a real system there will also be a thermal gradient in the x-direction, the direction of this gradient will vary based on the domain selected for analysis (e.g., center-line of laser, edge of melt pool). This may make the domain non-symmetric, and therefore, for simplicity an isothermal gradient in the x-direction is assumed. For the top surface of the domain, $y$ = 5 μm, the temperature is initialized to just below the melting point to induce solidification, $T_m - 0.1$ K (Fig. 1a). With the thermal gradient applied, the initial temperature at $y = 0$ is $T_m - 0.6$ K. At every model time step the temperature is updated while maintaining the same spatial temperature gradient across the y-direction.

*Selection of Mesh Size and Time Step:* The effects of altering the mesh size – $\Delta x$, $\Delta y$, and time step values – $\Delta t$, $\Delta t_{nuc}$, on the average grain size is compared against the reference case. Reducing the nucleation time step from 50 to 25 ns ($s_{dz} = 0.43\%$), halving the model time step from 10 to 5 ns ($s_{dz} = 1.3\%$), or reducing the mesh size from 40 to 20 nm ($s_{dz} = 2.7\%$) all result in grain sizes within the reference standard deviation. The model parameter timestep, $\Delta t = 10$ ns, selected is the largest value possible while simulations remain stable. Simulations are run for 20,000 timesteps (200 μs) for all cases, to determine steady-state grain size.

*Image Analyses:* A custom MATLAB grain area analysis tool was used to count the grain number and grain areas of simulated microstructures. A separate tool was developed to determine the fraction of nanoparticles on grain boundaries. Details of these programs are in Appendix III.

## 2.7 Strength vs Cost Model

Ultimately, the goal of adding nanoparticles is to improve the mechanical properties of the parent alloy at the cost of processing and manufacturing these new materials. Therefore, we present a comparison of the impact of nanoparticle area fraction on yield strength and cost. For submicron size grains, the two main strength additions of the nanoparticles will be grain boundary strengthening and Orowan bowing. Standard formulas are limited to random distributions of particles, so strength analysis is only applied to this distribution. Eq. (13) shows grain boundary strengthening

$$\sigma_{\text{HP}} = \frac{K_{\text{HP}}}{\sqrt{d_{\text{z}}}}. \tag{13}$$

$K_{\text{HP}}$ is a material specific Hall-Petch constant, and $d_{\text{z}}$ is the model predicted average grain diameter. $K_{\text{HP}}$ is assumed to be 740 MPa/$(\mu m)^{0.5}$, based on reported values for mild steel [52]. The second strengthening mechanism is Orowan bowing, which allows the nanoparticles themselves to impede dislocations. Eq. (14) gives the standard formula for Orowan Bowing [7]

$$\sigma_{\text{Or}} = \frac{0.4}{\pi} \frac{M}{\sqrt{1-v}} \frac{Gb}{L} \ln\left(\sqrt{\frac{2}{3}} \frac{d_{\text{z}}}{b}\right). \tag{14}$$

$M$ is the Taylor Factor, $v$ is Poisson's Ratio, $G$ is the Shear Modulus, $b$ is the Burgers vector, and $L$ is the average interparticle spacing for a random distribution of particles. Using the properties of SS316L, $M$ = 3.06, $v$ = 0.25, $G$ = 79 GPa, and $b$ = 0.254 nm [7]. $L$ is given by Eq. (15)

$$L = \sqrt{\frac{2}{3}} d_{\text{z}} \sqrt{\frac{\pi}{4} \frac{1}{\phi_{\text{np}}} - 1}. \tag{15}$$

The total alloy strength, $\sigma_{\text{y}}$, is given by Eq. (16) [7]

$$\sigma_{\text{y}} = \sigma_0 + \sqrt{\sigma_{\text{HP}}^2 + \sigma_{\text{Or}}^2}. \tag{16}$$

For SS316L, $\sigma_0$ = 592 MPa at room temperature and is the particle free alloy yield strength [7]. The cost analysis is conducted by considering the cost of powders for the base alloy, SS316L, and $Y_2O_3$ nanopowders. While $TiB_2$ particles are assumed for the nucleation model, $Y_2O_3$ has been used in commercial nanoparticle reinforced for decades allowing for more reasonable cost estimates [10]. The total material cost of the nanoparticle-modified alloy is dictated by the mass fraction, which linearly depends on the area fraction of the nanoparticles (Eq. (17))

$$Cost = (1 - \phi_{\text{np}})P_{\text{SS316L}} + \phi_{\text{np}}P_{\text{np}}. \tag{17}$$

SS316L powder cost is $P_{\text{SS316L}}$ = \$70/kg (3dpowderhub, Ontario, Canada, 25 μm) [53]. The baseline $Y_2O_3$ powder cost is $P_{\text{np}}$ = \$668/kg (US-nano, Texas, USA, 30-45 nm) [54]. With Eqs. (16–17) we calculate a mechano-economic metric, $\Theta$, which is the ratio of total alloy strength to total alloy cost, $\Theta = \sigma_y / Cost$.

## 3. Results and Discussion

Since this study considers many unique cases (70), select combinations are shown to illustrate takeaways more clearly. Trends for other cases are also discussed with videos of select cases shown in the Supplementary Materials. Videos are shown for the random, $\phi_{\text{np}}$ = 0.01 and $10^{-1}N_0$ case (Video S1) and clustered, $\phi_{\text{np}}$ = 0.01 and $10^3 N_0$ (Video S2).

### 3.1 Impact of Clustering and Nucleation Site Density on Solidification

Fig. 2 shows effects of nanoparticle clustering and nucleation site density on the transient evolution of the number of grains and their microstructures for a representative nanoparticle area fraction of $\phi_{\text{np}}$ = 0.05 compared against the particle-free case, i.e., $\phi_{\text{np}}$ = 0. Grain number versus time is shown for nucleation site densities of $0.1N_0$ (Fig. 2a) and $10N_0$ (Fig. 2e). Corresponding to the nucleation densities, 2D maps of the grain microstructures are shown only for the clustered distribution of particles — $0.1N_0$ (Fig. 2b–d) and $10N_0$ (Fig. 2f–h). Similar trends are obtained for other area fractions.

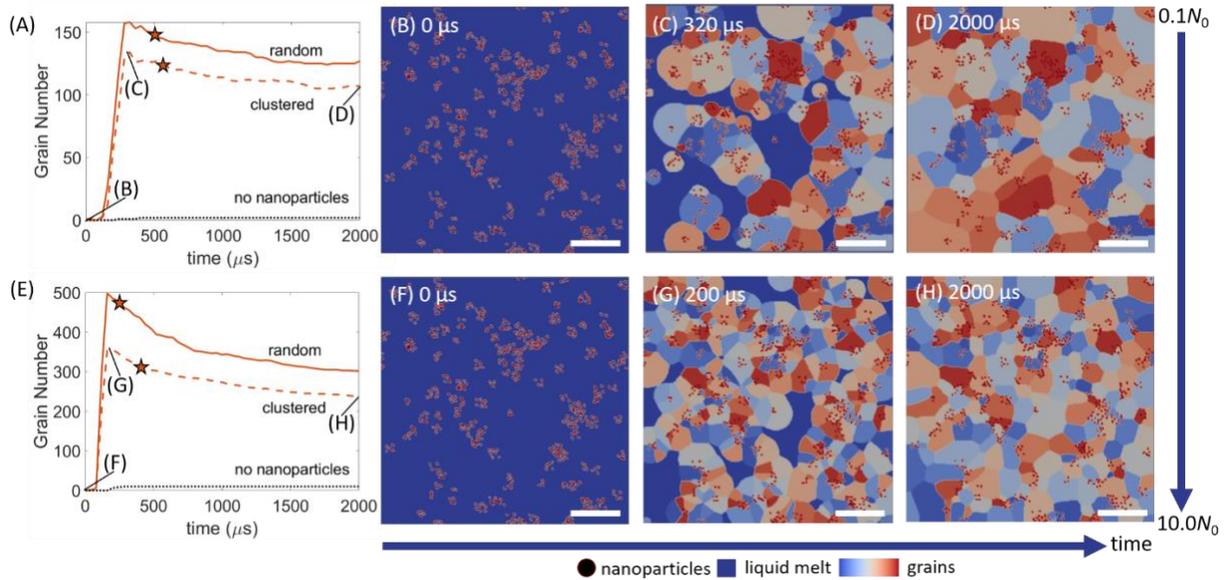

**Fig. 2.** Time evolution of grain microstructure for random and clustered distributions of nanoparticles with an area fraction of $\phi_{np}$ = 0.05 and for the case of no nanoparticles for nucleation site densities of (a–d) $0.1N_0$ and (e–h) $10N_0$. Stars in (a) and (e) represent the instance when solidification is complete. Snapshots of microstructures for the clustered distribution of nanoparticles at the (b,f) start of simulation, (c,g) peak number of grains, and the (d,h) final microstructure. Substrate is present at the bottom of the 2D domain and not shown in the figure. $N_0$ =1 corresponds to $1.95 \times 10^9$ nuclei/m$^2$. Scale bars are 1 μm.

For all cases modeled, there is an initial increase in the number of grains up to a maximum value beyond which it decreases. The maximum number of grains occurs before all the liquid melt is fully solidified (Fig. 2a,e). Through a competitive process, smaller grains tend to be consumed by larger grains resulting in grain coarsening [37]. After solidification is complete, grain coarsening continues to further reduce the grain number until a steady-state grain density, and equivalently, a grain size is attained.

The grain density is always much larger for any spatial distribution and concentration of nanoparticles in the alloy as compared to the nanoparticle free case (Fig. 2a). This is because heterogenous nucleation is more dominant on the nanoparticles than the substrate wall due to a larger nucleation surface area (per unit volume of the domain) for the nanoparticles. For $\phi_{np}$ = 0.05, the nanoparticle nucleation zone area is 15 times that of the substrate wall. For a fixed nucleation rate, a larger peak and steady-state number of grains is obtained for the random as compared to the clustered distribution of particles. As grains nucleate within a cluster zone, they quickly cover up nearby nanoparticles and therefore reduce the effective number of nucleation sites available, resulting in a lower number of grains. The final number of grains increases at a slower rate with site density as compared to the peak value due to grain coarsening. For the case of $0.1N_0$ and random distribution of nanoparticles, about 17% of grains are consumed from the peak value (Fig. 2a,c,d). When the nucleation site density increases to $10N_0$ a higher peak grain number is reached and a higher fraction of grains, 40%, are consumed through coarsening (Fig. 2e,g,h). This indicates that even in the presence of particle pinning a significant portion of grain nuclei will be consumed early into an alloy's lifetime. Additionally, as the nucleation rate increases from $0.1N_0$ to $10N_0$, the alloy solidifies sooner, 400 μs and 240 μs respectively, for the random particle distribution. This is because more grains are nucleated in the same domain, so solidifying grains reach pockets of liquid melt sooner.

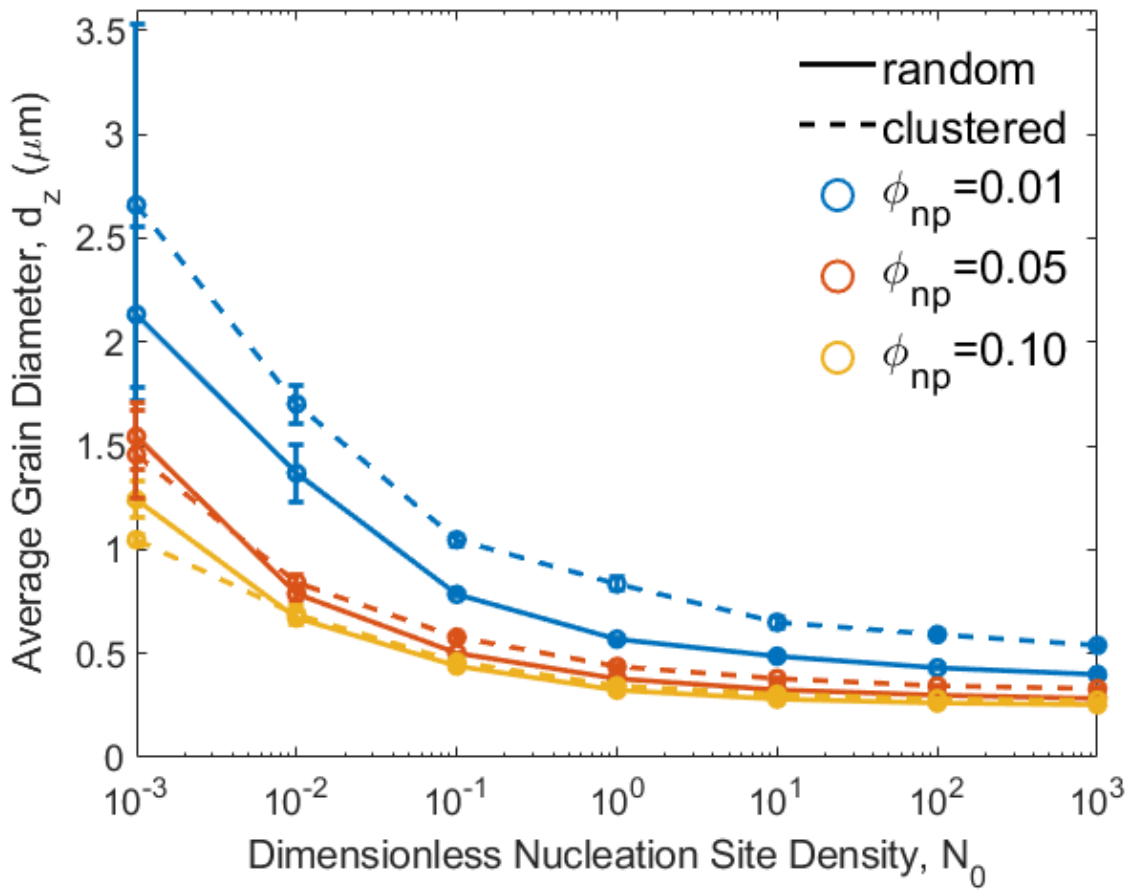

**Fig. 3.** Steady-state average grain size for random and clustered distributions as a function of nucleation site density, $N_0$, for various area fractions, $\phi_{np}$. Points shown are averages of 3 simulations per nucleation site density modeled and the error bars represent standard deviation. $N_0 = 1$ corresponds to $1.95 \times 10^9$ nuclei/m².

Fig. 3 shows the steady-state average grain size for modeled combinations of the nanoparticle concentrations ($\phi_{np} = 0.01$, 0.05, 0.10), spatial distributions (random and clustered), and nucleation site densities ($10^{-3}N_0$–$10^3N_0$). There is a relatively larger stochastic uncertainty in the average grain size for the low nucleation site densities. This is attributed to fewer, one or two dozen, grains that are nucleated as compared to hundreds of grains for larger nucleation site densities. For all spatial distributions and site densities, when area fraction increases, more particles nucleate new grains and pin grain boundaries resulting in smaller final grain sizes.

Clustering has a larger impact on increasing the final grain size relative to an ideal, random distribution when the area fraction is small. For example, for $\phi_{np}$ = 0.01 and $10^{-1}N_0$, clustering results in a 40% larger grain size than for the random distribution case. However, for a larger nanoparticle concentration of $\phi_{np}$ = 0.1, the clustered and random distributions converge and lead to nearly the same average grain size. At larger area fractions, many distinct clusters are present and nearly approximate a random distribution, resulting in similar grain sizes for both scenarios. This indicates that controlling processing/manufacturing parameters to limit clustering of nanoparticles is more important when the nanoparticle concentration is small (< 5% by area or volume).

At low nucleation site densities ($10^{-3}N_0$–$10N_0$), increases in site density (and corresponding nucleation rate) result in a significant decrease in final grain size for both random and clustered distributions. While grain size still decreases, increases in site density beyond $10N_0$ have diminishing impact on reducing the final grain size. This site saturation phenomenon is multifaceted and is discussed in more detail subsequently.

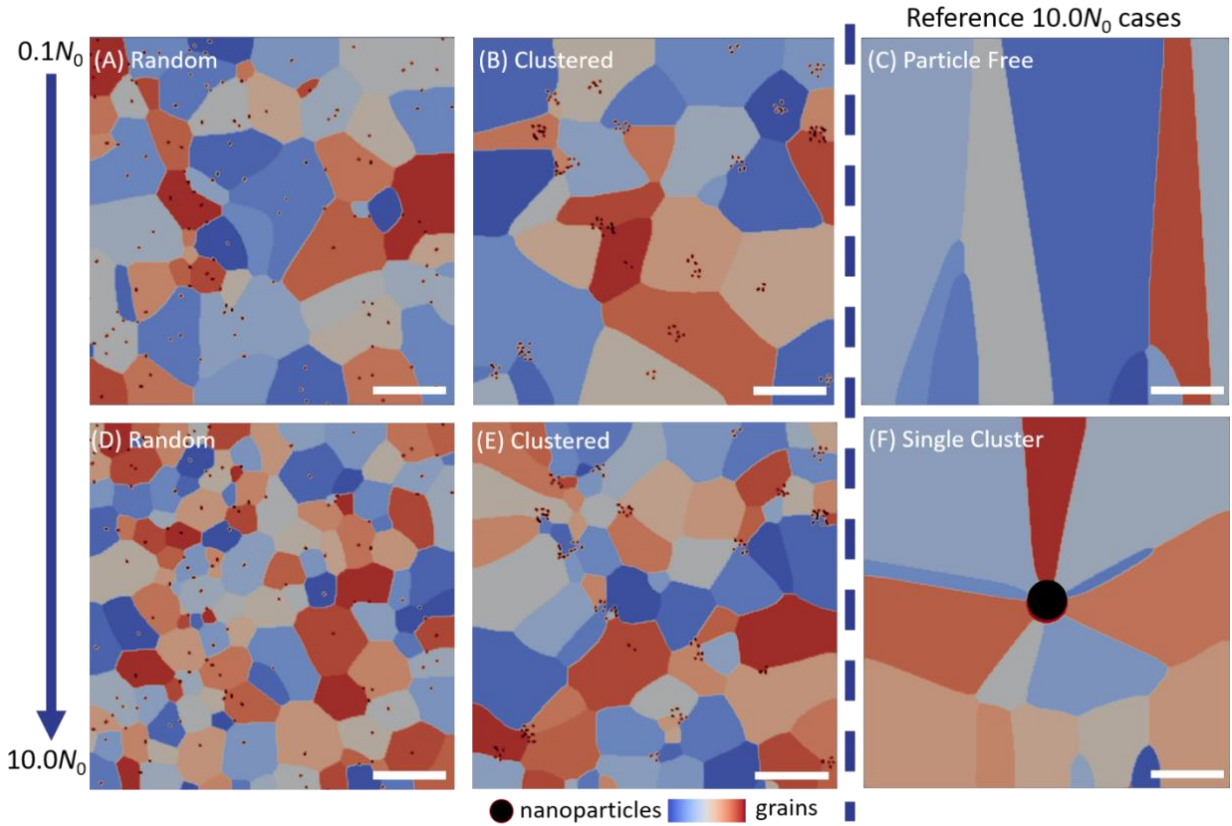

**Fig. 4.** Final grain microstructures after 2000 μs for select cases of 0.01 area fraction for random distributions with (a) $0.1N_0$, (d) $10N_0$; for clustered distributions with (b) $0.1N_0$, (e) $10N_0$. Reference cases shown with $10N_0$ for (c) particle free case and (f) with single large cluster of nanoparticles with 0.01 area fraction. Substrate is at bottom of domain but not shown. $N_0 = 1$ corresponds to $1.95 \times 10^9$ nuclei/m². Scale bars are 1 μm.

Fig. 4 provides further insight into the effects of nanoparticle distributions and nucleation site density on the final grain microstructures for an area fraction of 0.01 and for nucleation site densities of $0.1N_0$ and $10N_0$. For comparison, final microstructures are shown for the extreme cases with no nanoparticles (Fig. 4c) and one single large cluster with an area fraction of 0.01 and $10N_0$ (Fig. 4f). When comparing the impact of increasing site density on microstructure for both the random (Fig. 4a,d) and clustered (Fig. 4b,e) cases, a larger site density results in a more refined microstructure as expected. However, while the random distribution case has a spatially homogenous distribution of grains, the clustered case is marked by regions of small grains and regions of large grains. Smaller grains tend to appear near regions with high cluster concentrations, and larger grains grow away from clusters into particle free regions. For the particle free case (Fig. 4c), since all grains must nucleate from the substrate, long, columnar grains grow with smaller, shorter grains slowly being consumed by longer grains. In contrast, Fig. 4a,b,d,e demonstrate that even 0.01 area fraction of particles are effective in supporting the columnar-to-equiaxed grain transition [30]. The single cluster case, representing the most severe degree of clustering possible, shows grains grow in a sunburst style shape away from the cluster in the center of the domain (Fig 4f). Since columnar grains on the edges of the substrate are farther from the cluster, they can grow longer than grains from the middle of the substrate. This shows that when the average distance between two nucleation sites increases, grains growing between those points can become larger/longer.

To further quantify the final microstructures, Fig. 5 shows the impact of area fraction, nucleation site density, and clustering on grain size distributions. Skewness represents the extent of deviation from a normal distribution; therefore, a normal distribution will have a skewness of 0 [55]. For an ideal, homogenous microstructure, most grains will be close to the average size with a relatively larger fraction of grains smaller than the average to avoid a few larger grains from dominating the microstructure. This would result in a positively skewed distribution with skewness of 0.5–1.0 [18,56]. A random distribution of nanoparticles with $\phi_{np} = 0.01$ and $10N_0$ sites, with an average grain size of 487 nm, standard deviation of 11 nm and a skewness of 0.575, is selected as a reference case for Fig. 5a–c. When the nucleation site density is held fixed ($10N_0$) and the area fraction increases to 0.1, the skewness increases to 0.716 and standard deviation decreases, indicating a higher fraction of grains are smaller than the average compared to reference case. When the area fraction is fixed ($\phi_{np}$ = 0.01) and the nucleation site density is decreased to $0.1N_0$ (Fig. 5b), the distribution and standard deviation widen (skewness = 0.386) with a higher likelihood of having larger grains (> 1000 μm) than the reference case. This indicates the nucleation rate, controlled by site density, impacts grain distribution. When nucleation rate ($10N_0$) and area fraction ($\phi_{np}$=0.01) are fixed and the distribution type changes from random to clustered, the size distribution and standard deviation again widen with a skewness value of 0.325 (Fig. 5c). As Fig. 4e shows, this wider grain size distribution is marked by small grains near concentrations of nanoparticle clusters and large grains far away from clusters.

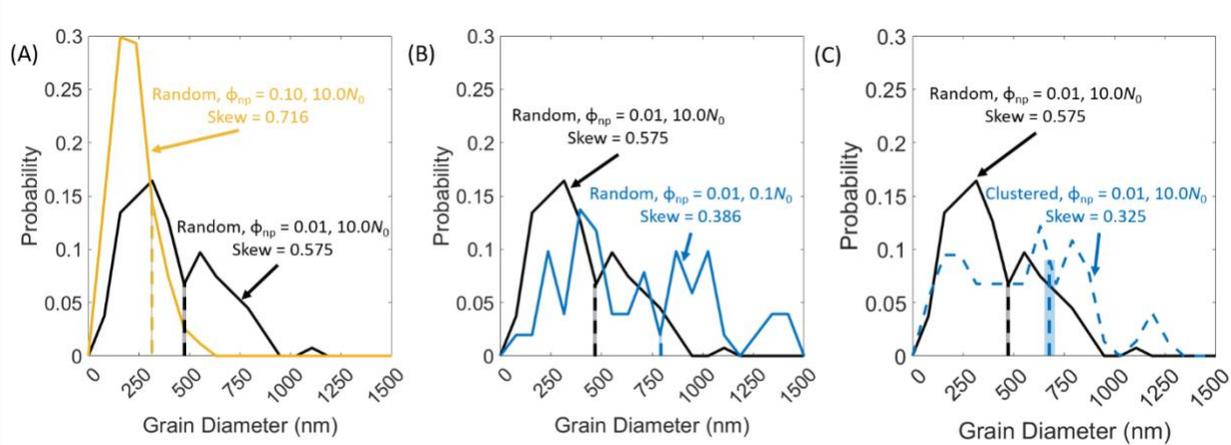

**Fig. 5.** Comparison of final grain size distribution to reference case of random with area fraction, $\phi_{np}$ = 0.01, and nucleation site density, $10N_0$. Impact of (a) area fraction: random, $\phi_{np}$ = 0.10 and $10N_0$, (b) nucleation rate: random, $\phi_{np}$ = 0.01 and $0.1N_0$, (c) clustering: clustered, $\phi_{np}$ = 0.01 and $10N_0$. Dotted vertical lines indicate the average grain size for each case with shaded regions indicating standard deviation. $N_0$ = 1 corresponds to $1.95 \times 10^9$ nuclei/m$^2$.

For all the cases shown in Fig. 5, the largest grain size only approaches twice the average grain size. While cutoffs can vary, abnormally large grains, which can induce detrimental mechanical properties, are typically at least 5–10x the average grain size [1,2,57,58]. All simulations in our study are in the normal grain growth regime (even for $\phi_{np}$ = 0.01, $10^{-3}N_0$), likely because of the assumption of perfect pinning by nanoparticles [56].

### 3.3 Zener Grain Size Depends on Nucleation Rate

The results of Fig. 3 can be analyzed through the Zener Equation, which provides an estimation of the average grain size as a function of area fraction without having to conduct experiments for intermediate area fractions. Traditionally, the limiting average grain diameter, $d_z$, of nanoparticle reinforced alloys is given by the generic Zener diameter (Eq. (18)),

$$d_z = 2r_p \frac{K}{(\phi_{np})^m} \qquad (18)$$

where $d_z$ represents an estimate for grain size, and $r_p$ is the nanoparticle radius [26]. Eq. (18) is applicable to both 2D and 3D simulations/experiments [59]. For the classic Zener equation, which was determined for a random distribution of nanoparticles and grains, $K$ = 4/3 and $m$ = 1, but these values typically overestimate grain size [59]. Physically, the constant $K$ is influenced by the particle-grain pinning force [56,59], and $m$ describes how changes in area fraction influence grain size [57]. Fig. 6a shows the predicted grain size versus area fraction for bounding cases of $10^{-3}N_0$ and $10^3N_0$ for random and clustered cases. Two reference cases are shown in Fig. 6a, including the classic Zener Equation ($K$ = 4/3 and $m$ = 1) [58], and 2D Monte Carlo simulations of grain recrystallization with a random distribution of pinning nanoparticles ($K$ = 1.7 and $m$ = 1/2) [57]. For $10^{-3}N_0$, grain sizes match the classic Zener Equation at $\phi_{np}$ = 0.05, but dramatically different for other area fractions. When the site density increases to $10^3N_0$, both distributions lead to average grain sizes below the classic Zener equation. While for low area fractions the clustered grain size is slightly larger, both distributions converge at an area fraction of 0.1 with grain sizes reaching 300 nm approaching the Srolovitz et. al. curve [57].

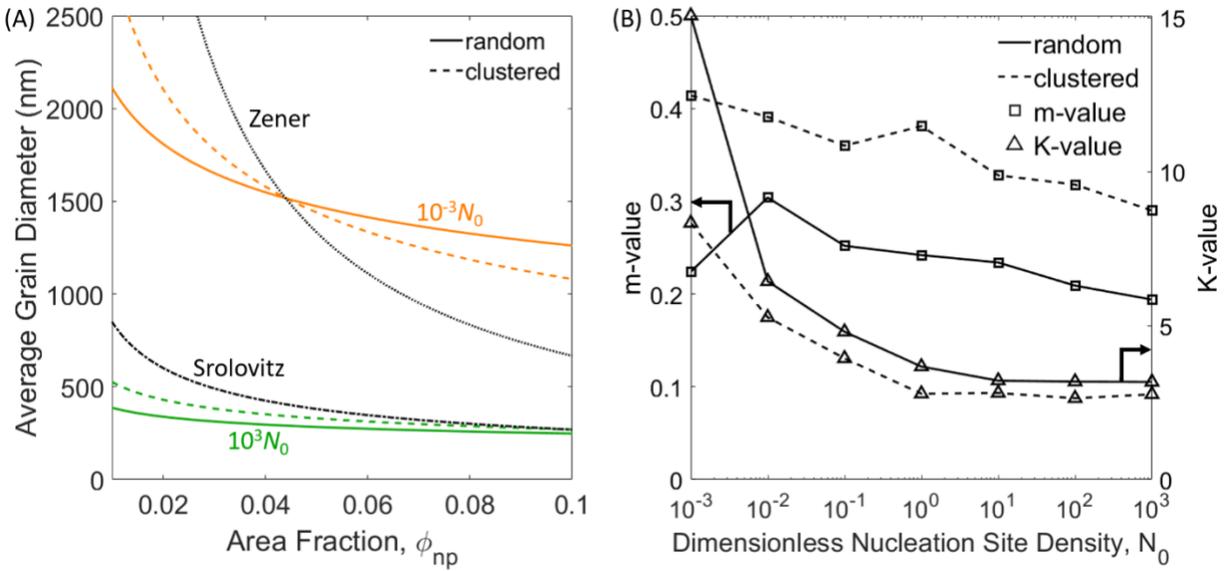

**Fig. 6.** Modified Zener Diameters for random and clustered distributions. (a) Grain diameter versus area fraction for nucleation site densities of $10^{-3}N_0$ and $10^3N_0$. All other $N_0$ will fit between these two extremes. Reference fits are from the original Zener Equation as recorded by Smith [58] and Srolovitz et. al. [57]. (b) Impact of nucleation site density on $m$ and $K$, which are fitting constants in the Zener Equation (Eq. (13)). $N_0 = 1$ corresponds to $1.95 \times 10^9$ nuclei/m$^2$.

Fig. 6b shows the best-fit values for $K$ and $m$ as a function of spatial distributions and nucleation site densities using the data from Fig. 3. Consistent with the clustered grain size being more sensitive to area fraction (Fig. 6a), for each $N_0$ modeled, the $m$ values of the clustered (0.299–0.407) are higher than for the random distribution (0.195–0.289). The $m$ values are relatively constant with a slight decrease as nucleation site density increases. In contrast, the $K$ value decreases, 4.7 times for the random and 3.0 times for the clustered cases, when site density increases from $10^{-3}N_0$ to $10N_0$. Beyond $10N_0$, $K$ values level off. Table 2 presents modified Zener

Equations that account for nucleation site density dependency of $K$ while assuming a constant, average $m$ value for each distribution.

Most previous 2D simulations, which ignore heterogenous nucleation, have $m$ values between 0.3–0.5, whereas our study has lower $m$ values between 0.195–0.407 [59]. Our study suggests nanoparticles serving as nucleation sites may lower the $m$ value with this effect reduced for clustered cases. While the traditional Zener equation (Eq. (18)) has no nucleation rate dependence, Avrami analysis for solidification dynamics predicts an inverse relationship between the final grain size and the nucleation rate, dictated by the degrees of freedom of the system [60]. Our results also predict that grain size is inversely related to nucleation rate. Although a full-fledged Avrami analysis is outside the scope of this study, our analysis predicts consistent trends for the effects of nucleation rate on the grain size.

Table 2. Nucleation rate dependent grain size predictions for random and clustered distributions fitted for nucleation site densities ($10^{-3}N_0$–$10^3N_0$) and area fractions ($\phi_{np}$ = 0.01–0.10). $N_0$ = 1 corresponds to $1.95 \times 10^9$ nuclei/m$^2$.

| Nanoparticle Distribution | Best Fit Equations | $R^2$ |
|---|---|---|
| Random | $d_z = 2r_p \dfrac{K}{\phi_{np}^{0.25}}, K = \dfrac{5.0}{N_0^{0.15}}$ | 0.972 |
| Clustered | $d_z = 2r_p \dfrac{K}{\phi_{np}^{0.33}}, K = \dfrac{4.0}{N_0^{0.15}}$ | 0.977 |

### 3.4 Grain Size Saturation is Explained by Particle Pinning Fraction

Since nanoparticles can only effectively prevent grain growth when they are on grain boundaries, the saturation in grain size for $> 10N_0$ (Fig. 6a) is explained by quantifying the fraction of nanoparticles on grain boundaries, $\Gamma$. Based on the classic Zener equation (Eq. (18)), which assumes no correlation between nanoparticle and grain boundary locations, the fraction of nanoparticles pinned on grain boundaries can be predicted using Eq. (19) [61,62]

$$\Gamma_Z = (3/2)\phi_{np}. \tag{19}$$

For area fractions of 0.01, 0.05, and 0.10, Eq. (19) predicts 1.5%, 7.5%, and 15%, respectively for the fraction of nanoparticles on grain boundaries. However, as Fig. 7a shows, even for the smallest nucleation site density modeled ($10^{-3}N_0$), the actual fraction pinned is much larger, 8–24%, than predicted from Zener's equation. The 2D microstructures and calculated pinned fractions indicate a strong preference of nanoparticles present on grain boundaries, due to grain boundary pinning. For all particle distributions and area fractions, as the nucleation site density increases ($10^{-3}N_0$–$10N_0$), the pinning fraction increases rapidly initially and asymptotes ($10^2$–$10^3N_0$) towards pinned fractions in the range of 0.85–0.90. Published phase-field and Monte Carlo studies of homogenous nucleation also found similar trends. For these studies, when the initial grain number increases, analogous in our study to increasing the nucleation rate, the pinning fraction increases before reaching a saturation value of 0.8–0.85 for 2D cases [41,61]. Experimental studies of a Ni-based superalloy also found a preference of nanoparticles at the grain boundaries again reaching a saturation value between 0.8-0.9 and yielded the smallest grain sizes [63].

Fig. 7b,c, which considers all 5 area fractions (0.01–0.1), reveal that as the pinning fraction plateaus the steady-state grain size does as well. Fig. 7b shows that for the random distribution of particles regardless of nucleation rate or area fraction, all grain sizes collapse to single curve with strong correlation ($R^2 = 0.980$). This simple analytical function leads us to propose that the

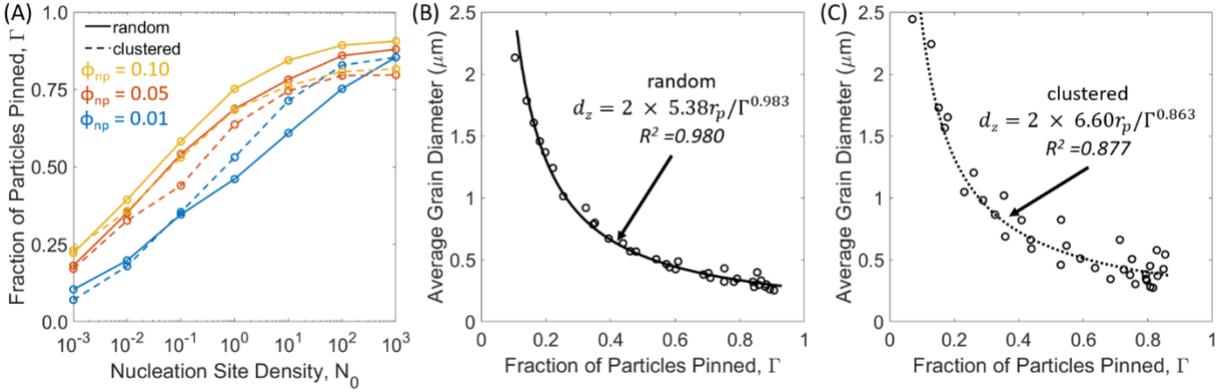

**Fig. 7.** (a) Impact of nucleation site density, $N_0$, on fraction of particles pinning grain boundaries, $\Gamma$, for select area fractions of 0.01, 0.05, 0.10. (b–c) Impact of pinning fraction on final grain diameter with best fit curves for all area fractions. (b) All random cases and (c) all clustered cases. $N_0 = 1$ corresponds to $1.95 \times 10^9$ nuclei/m$^2$.

fundamental governing mechanism behind Zener equations is the fraction of nanoparticles on grain boundaries, at least for the area fractions considered and when ideal pinning of nanoparticles is modeled. The reason area fraction and nucleation rate impact the final grain size is because they change the pinning fraction. Fig. 7c extends this analysis to a clustered distribution of particles. However, the clustered distribution does not result in as good a fit ($R^2 = 0.877$), particularly at higher fractions pinned. The pinning fraction required to achieve a fixed grain size is larger for the clustered compared to the random distributions of the nanoparticles, which is likely due to clustering allowing free grain growth away from cluster centers (Fig. 4e). These results show that the spatial distribution of the nanoparticles is an additional factor that can influence grain size even with the pinning fraction being a dominant factor. Overall, the strong correlation predicted between pinning fraction and grain size calls for researchers to report/extract pinning fraction data from microstructural characterization, in addition to grain size distributions, to determine if the nanoparticles are effective in reducing grain size.

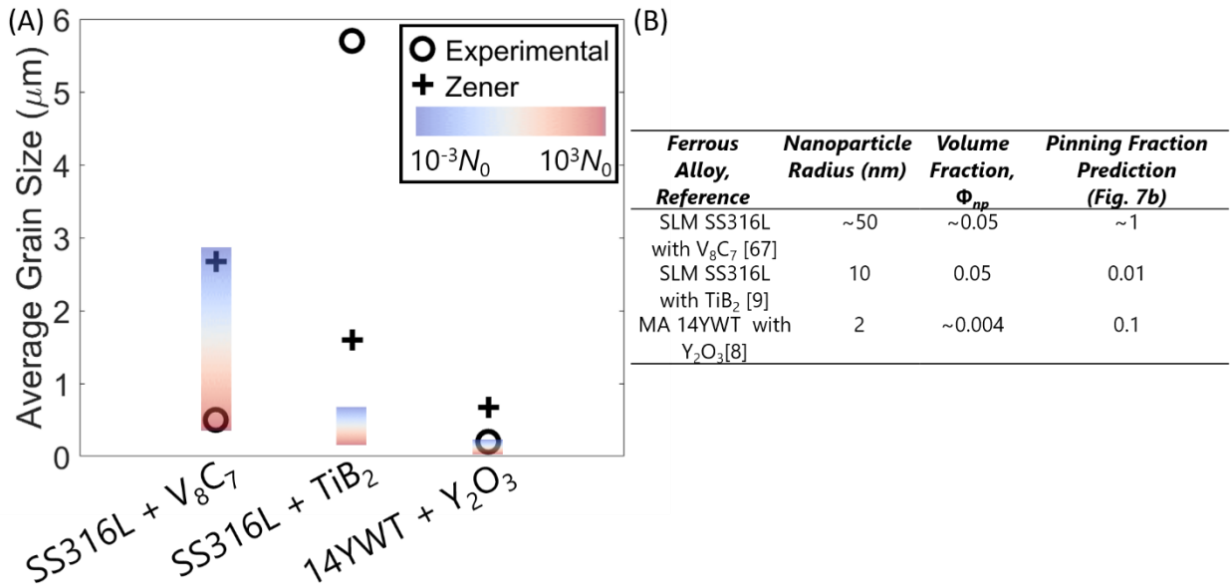

**Fig. 8.** (a) Predicted vs Experimental Grain Sizes reported in the literature for select nanoparticle reinforced alloys. Nucleation rate dependent grain size predictions come from the random distribution case of Table 2. (b) Table showing additional pertinent alloy information.

Fig. 8(a) shows a comparison of experimentally reported grain sizes and nanoparticle volume

fractions for select alloys with the Zener model (Eq. (18)) and the predicted range of grain size

(Table 2) corresponding to the nucleation rates considered in this study. For our predictions, we assumed a random distribution of nanoparticles for the sake of generality across different alloys.

The table in Fig. 8b lists the experimental nanoparticle radius, volume fraction, and pinning fraction prediction of Fig. 7b.

For all cases, the classic Zener Equation incorrectly predicts the grain sizes as it neglects the

effects of nucleation rate. For the SLM SS316L alloys reinforced with $V_8C_7$ nanoparticles, while

the classic Zener relationship overpredicts the experimentally measured grain size, our

predictions match the measured grain size for a nucleation density of $150N_0$ [64]. For the $TiB_2$

reinforced alloy, the experimental grain size is larger than the classic Zener Equation, which may be due to the limited ability of the TiB$_2$ nanoparticles to pin grain boundaries. This is backed by the relatively small lattice mismatch between SS316L and TiB$_2$ [65], which also indicates that the assumption of ideal particle pinning in our model predictions may not be reasonable for this case. Therefore, the predicted nucleation rate to match the experimental grain size is much lower and outside the fitted range of nucleation site densities. Such a large grain size also maps with rather small pinning fraction estimate, which is also strictly outside our prediction range. Although the phase-field model was developed specifically for rapid solidification processes, we also compare our predictions with mechanically alloyed (MA) ferrous 14YWT reinforced with Y$_2$O$_3$. The Y$_2$O$_3$ reinforced alloy has a grain size below the Zener limit [8] and the experimental grain size can be matched for $2 \times 10^{-3} N_0$ and with a 0.4% nanoparticle volume fraction, or equivalently with a pinning fraction of 0.1.

Ultimately, the benefit of our predictions is that it can facilitate insight on nucleation rates and pinning fractions from experimental measurements of grain size. In this comparison case, our predictions indicate that V$_8$C$_7$ particles can likely induce a higher nucleation rate than either the TiB$_2$ or Y$_2$O$_3$ particles. Our simulations also predict the V$_8$C$_7$ and Y$_2$O$_3$ cases, compared to TiB$_2$, should have a much higher fraction on nanoparticles on grain boundaries. This matches experimental observations and materials characterization measurements for these materials. For the MA Y$_2$O$_3$ reinforced alloy, while the classic Zener distribution, Eq. (19), predicts 0.006 fraction of nanoparticles to be on grain boundaries, our model predicts a 0.1 fraction. The reported fraction is roughly 0.25 for the Y$_2$O$_3$ nanoparticles on the grain boundaries [8]. Therefore, while our prediction is much closer than the Zener estimate, the discrepancies could be due to several factors including: (a) differences in pinning dynamics in 2D vs 3D, and (b) we assumed rigid nanoparticles, but microstructure analysis reveals evidence of particle pushing during solidification, which may influence the constants in the grain size prediction equations (Fig. 7b). Nonetheless, these results reinforce the importance of accounting for the pinning fraction to perform grain size analysis.

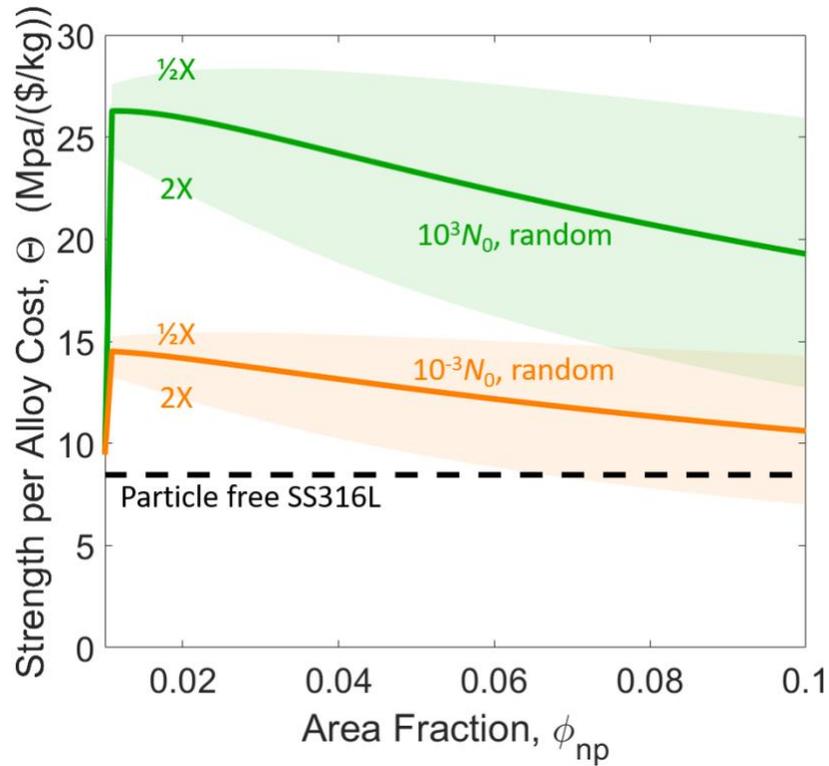

**Fig. 9** Impact of nanoparticle area fraction vs strength per unit cost of alloy. Select cases of a random distribution with nucleation site densities of $10^3 N_0$ and $10^{-3} N_0$. Solid lines represent analysis with baseline $Y_2O_3$ powder cost with shaded regions showing bounds of nanoparticle cost estimates. $N_0$ corresponds to $1.95 \times 10^9$ nuclei/m$^2$.

## 3.6 Economic Trade-offs for Nanoparticle Area Fraction

Fig. 9 shows the mechano-economic metric, $\Theta$, which is the total alloy strength per unit cost (MPa/($/kg)) as a function of area fraction (equivalently volume fraction in 3D) for the nanoparticle modified alloys. Since nanoparticle cost can vary widely based on material and powder batch size, a brief sensitivity analysis was conducted assuming the nanoparticle cost to

be half or twice the baseline value (Section 2.7). For comparison, the nanoparticle free case is also shown, which has a value of 8.4 MPa/($/kg) for this metric. Cases are shown for the random distribution of nanoparticles using the lower and upper grain size limits, i.e., $10^3 N_0$ and $10^{-3} N_0$. In most cases simulated, $\Theta$ is comparatively better than the nanoparticle free case, even for the lowest nucleation site density, which points to the promise of nanoparticle additions. For both distributions, $\Theta$ increases rapidly before decreasing, indicating an optimal area fraction close to $\phi_{np} = 0.01$–$0.02$. This is because the largest grain size reductions compared to the particle free case occur at low area fractions. Even when nanoparticle cost is doubled, the $10^3 N_0$ case has a higher $\Theta$ than the particle free case for all area fractions considered. However, for the $10^{-3} N_0$ and with twice the baseline cost of nanoparticles, $\Theta$ drops below the particle free case when area fraction approaches 0.1 indicating a critical cost beyond which nanoparticle addition will be increasingly less lucrative. A more in-depth mechano-economic model that accounts for materials processing costs and imperfect particle pinning will shift the optimal area fraction values. However, the projected trends still provide useful insight to guide selection of nanoparticle area fractions (and therefore weight percent) to strengthen metal alloys cost effectively.

## 4. Conclusion

In this study, we develop and apply a phase-field model to simulate alloy solidification and grain growth with a dispersion of nanoparticles, motivated by their potential to improve yield strength. During many laser additive manufacturing processes, nanoparticle clusters will form due to large thermal gradients. A direct advancement of this model is an explicit study of particle clustering on microstructure. The model investigated the effects of nanoparticle distribution – random and clustered spatial distributions, area fraction (0.01–0.10), and nucleation site density ($1.95 \times 10^6$–$1.95 \times 10^{12}$ nuclei/m$^2$) on grain growth dynamics. The model incorporates effects of nanoparticles on heterogenous nucleation and grain boundary pinning.

We show for low area fractions, ~0.01, clustering of nanoparticles results in 15-45% larger grain diameters than the random case. The clustered cases also have a wider, more heterogenous grain size distribution. As the area fraction increases the random and clustered grain sizes converge showing that clustering's impact is more pronounced at lower area/volume fractions of nanoparticles. While the classic Zener Equation ignores the impact of nucleation rate, we developed a modified Zener Equation to predict grain size as a function of nucleation rate. Specifically, our results demonstrate that grain size is inversely related to nucleation site density as $1/(N_{\text{sites}})^{0.15}$ for both random and clustered 2D distributions. As the nucleation rate increases the grain size dramatically reduces before it asymptotes when the grain size is 3–5 times smaller than the low nucleation rate case.

Our predictions also underscore that the grain size for various nanoparticle volume fractions and nucleation rates are primarily governed by the fraction of particles located on grain boundaries, i.e., the pinned fraction of particles. This dependence is moderately influenced by the spatial distributions of the particles. We deduced governing relationships between grain size and pinning fraction of $d_z = 2 \times 5.38 r_p/\Gamma^{0.983}$ ($R^2 = 0.980$) and $d_z = 2 \times 6.60 r_p/\Gamma^{0.863}$ ($R^2 = 0.877$) for the random and clustered distributions modeled. These results demonstrate that while the grain size dependence on pinning fraction is dominant for the random distribution of nanoparticles, the spatial heterogeneity introduced by clustering makes grain size predictions more complex.

Our model predictions when compared to experiments reveal that nucleation rate and pinning fraction dependence on grain size are crucial factors and can explain why many experimental data deviate from the classic Zener Equation predictions. A mechano-economic metric was developed to compare the strength and cost advantages of alloys reinforced with nanoparticles compared to the particle-free case. Addition of nanoparticles improved the alloy strength per unit cost even at high area fractions close to 0.10 when reported materials costs were considered for $Y_2O_3$ strengthened SS316L. The optimum strength to cost ratio occurs between 0.01–0.02 area fraction of nanoparticles and suggests that the largest strength gains on a cost basis occur at low area fractions.

Overall, this study provides a new alloy solidification model, and reveals that grain size is a function of particle clustering and nucleation rate. Model results offer valuable insight to guide the development of next generation, stronger nanoparticle reinforced alloys.

## Acknowledgements

The information, data, or work presented herein was funded in part by the Advanced Research Projects Agency-Energy (ARPA-E), U.S. Department of Energy, under Award Number DE-AR0001123 under co-operative agreement with Michigan State University and the University of Michigan. The views and opinions of authors expressed herein do not necessarily state or reflect those of the United States Government or any agency thereof. Bala Chandran acknowledges startup funding from the College of Engineering and the Department of Mechanical Engineering at the University of Michigan. This research was supported in part through computational resources and services provided by Advanced Research Computing (ARC), a division of Information and Technology Services (ITS) at the University of Michigan, Ann Arbor. Authors acknowledge colleagues at the University of Michigan – Luisa Barrera for sharing a MATLAB script to perform grain size analysis, Aditya Sundar for insightful conversations, and David Montiel for technical assistance on the PRISMS-PF software.

# References

[1]     W.H. Yu, S.L. Sing, C.K. Chua, C.N. Kuo, X.L. Tian, Particle-reinforced metal matrix nanocomposites fabricated by selective laser melting: A state of the art review, Progress in Materials Science. 104 (2019) 330–379. https://doi.org/10.1016/j.pmatsci.2019.04.006.

[2]     N. Oono, Q.X. Tang, S. Ukai, Oxide particle refinement in Ni-based ODS alloy, Materials Science and Engineering A. 649 (2016) 250–253. https://doi.org/10.1016/j.msea.2015.09.094.

[3]     Q. Tang, T. Hoshino, S. Ukai, B. Leng, S. Hayashi, Y. Wang, Refinement of Oxide Particles by Addition of Hf in Ni-0.5 mass%Al-1 mass%Y 2 O 3 Alloys, Materials Transactions. 51 (2010) 2019–2024. https://doi.org/10.2320/matertrans.M2010163.

[4]     D.T. Hoelzer, J. Bentley, M.A. Sokolov, M.K. Miller, G.R. Odette, M.J. Alinger, Influence of particle dispersions on the high-temperature strength of ferritic alloys, Journal of Nuclear Materials. 367-370 A (2007) 166–172. https://doi.org/10.1016/j.jnucmat.2007.03.151.

[5]     C.M. Cepeda-Jiménez, M.T. Pérez-Prado, Processing of nanoparticulate metal matrix composites, in: Comprehensive Composite Materials II, Elsevier, 2017: pp. 313–330. https://doi.org/10.1016/B978-0-12-803581-8.09984-7.

[6]     B.W. Baker, T.R. McNelley, L.N. Brewer, Grain size and particle dispersion effects on the tensile behavior of friction stir welded MA956 oxide dispersion strengthened steel from low to elevated temperatures, Materials Science and Engineering A. 589 (2014) 217–227. https://doi.org/10.1016/j.msea.2013.09.092.

[7]     B. AlMangour, Y.K. Kim, D. Grzesiak, K.A. Lee, Novel TiB2-reinforced 316L stainless steel nanocomposites with excellent room- and high-temperature yield strength developed by additive manufacturing, Composites Part B: Engineering. 156 (2019) 51–63. https://doi.org/10.1016/J.COMPOSITESB.2018.07.050.

[8]     J.H. Kim, T.S. Byun, D.T. Hoelzer, S.W. Kim, B.H. Lee, Temperature dependence of strengthening mechanisms in the nanostructured ferritic alloy 14YWT: Part I-Mechanical and microstructural observations, Materials Science and Engineering A. 559 (2013) 101–110. https://doi.org/10.1016/j.msea.2012.08.042.

[9]     J.H. Kim, T.S. Byun, D.T. Hoelzer, C.H. Park, J.T. Yeom, J.K. Hong, Temperature dependence of strengthening mechanisms in the nanostructured ferritic alloy 14YWT: Part II-Mechanistic models and predictions, Materials Science and Engineering A. 559 (2013) 111–118. https://doi.org/10.1016/j.msea.2012.08.041.

[10]    G.B. Schaffer, M.H. Loretto, R.E. Smallman, J.W. Brooks, The stability of the oxide dispersion in INCONEL alloy MA6000, Acta Metallurgica. 37 (1989) 2551–2558. https://doi.org/10.1016/0001-6160(89)90053-9.

[11]    R. Cueff, H. Buscail, E. Caudron, C. Issartel, F. Riffard, Oxidation behaviour of Kanthal APM and Kanthal AF at 1173 K: Effect of yttrium alloying addition, Surface Engineering. 19 (2003) 58–64. https://doi.org/10.1179/026708403225002469.

[12]    T.C. Lin, C. Cao, M. Sokoluk, L. Jiang, X. Wang, J.M. Schoenung, E.J. Lavernia, X. Li, Aluminum with dispersed nanoparticles by laser additive manufacturing, Nature Communications. 10 (2019). https://doi.org/10.1038/S41467-019-12047-2.

[13]    D. Gu, H. Zhang, D. Dai, M. Xia, C. Hong, R. Poprawe, Laser additive manufacturing of nano-TiC reinforced Ni-based nanocomposites with tailored microstructure and performance, Composites Part B: Engineering. 163 (2019) 585–597. https://doi.org/10.1016/j.compositesb.2018.12.146.

[14]    J.P. Kruth, L. Froyen, J. van Vaerenbergh, P. Mercelis, M. Rombouts, B. Lauwers, Selective laser melting of iron-based powder, in: Journal of Materials Processing Technology, 2004: pp. 616–622. https://doi.org/10.1016/j.jmatprotec.2003.11.051.

[15]    T. Kinoshita, M. Ohno, Phase-field simulation of abnormal grain growth during carburization in Nb-added steel, Computational Materials Science. 177 (2020) 109558. https://doi.org/10.1016/j.commatsci.2020.109558.

[16]    D. Gu, H. Wang, F. Chang, D. Dai, P. Yuan, Y.C. Hagedorn, W. Meiners, Selective laser melting additive manufacturing of TiC/AlSi10Mg bulk-form nanocomposites with tailored microstructures and properties, in: Physics Procedia, Elsevier B.V., 2014: pp. 108–116. https://doi.org/10.1016/j.phpro.2014.08.153.

[17]    Q. Tan, J. Zhang, N. Mo, Z. Fan, Y. Yin, M. Bermingham, Y. Liu, H. Huang, M.X. Zhang, A novel method to 3D-print fine-grained AlSi10Mg alloy with isotropic properties via inoculation with LaB6 nanoparticles, Additive Manufacturing. 32 (2020). https://doi.org/10.1016/J.ADDMA.2019.101034.

[18]    X. Li, W. Tan, Numerical investigation of effects of nucleation mechanisms on grain structure in metal additive manufacturing, Computational Materials Science. 153 (2018) 159–169. https://doi.org/10.1016/j.commatsci.2018.06.019.

[19]    M. Balog, P. Krizik, O. Bajana, T. Hu, H. Yang, J.M. Schoenung, E.J. Lavernia, Influence of grain boundaries with dispersed nanoscale Al2O3 particles on the strength of Al for a wide range of homologous temperatures, Journal of Alloys and Compounds. 772 (2019) 472–481. https://doi.org/10.1016/j.jallcom.2018.09.164.

[20]    J.Q. Xu, L.Y. Chen, H. Choi, X.C. Li, Theoretical study and pathways for nanoparticle capture during solidification of metal melt, J. Phys.: Condens. Matter. 24 (2012) 255304–255314. https://doi.org/10.1088/0953-8984/24/25/255304.

[21]    M.F. Ashby, R.M.A. Centamore, The Dragging of Small Oxide Particles by Migrating Grain Boundaries in Copper, Acta Metallurgica. 16 (1968) 1081–1092.

[22]    M. Nganbe, M. Heilmaier, Modelling of particle strengthening in the γ' and oxide dispersion strengthened nickel-base superalloy PM3030, Materials Science and Engineering A. 387–389 (2004) 609–612. https://doi.org/10.1016/j.msea.2004.01.109.

[23]    Y. Zhang, Y. Yu, L. Wang, Y. Li, F. Lin, W. Yan, Dispersion of reinforcing micro-particles in the powder bed fusion additive manufacturing of metal matrix composites, Acta Materialia. 235 (2022) 118086. https://doi.org/10.1016/J.ACTAMAT.2022.118086.

[24] H. Aufgebauer, J. Kundin, H. Emmerich, M. Azizi, C. Reimann, J. Friedrich, T. Jauß, T. Sorgenfrei, A. Cröll, Phase-field simulations of particle capture during the directional solidification of silicon, Journal of Crystal Growth. 446 (2016) 12–26. https://doi.org/10.1016/j.jcrysgro.2016.04.032.

[25] Y. Yang, C. Doñate-Buendía, T.D. Oyedeji, B. Gökce, B.X. Xu, Nanoparticle tracing during laser powder bed fusion of oxide dispersion strengthened steels, Materials. 14 (2021). https://doi.org/10.3390/ma14133463.

[26] C. Schwarze, R. Darvishi Kamachali, I. Steinbach, Phase-field study of zener drag and pinning of cylindrical particles in polycrystalline materials, Acta Materialia. 106 (2016) 59–65. https://doi.org/10.1016/j.actamat.2015.10.045.

[27] N. Moelans, B. Blanpain, P. Wollants, A phase field model for the simulation of grain growth in materials containing finely dispersed incoherent second-phase particles, Acta Materialia. 53 (2005) 1771–1781. https://doi.org/10.1016/j.actamat.2004.12.026.

[28] Y. Suwa, Y. Saito, H. Onodera, Phase field simulation of grain growth in three dimensional system containing finely dispersed second-phase particles, Scripta Materialia. 55 (2006) 407–410. https://doi.org/10.1016/j.scriptamat.2006.03.034.

[29] L. Gránásy, F. Podmaniczky, G.I. Tóth, G. Tegze, T. Pusztai, Heterogeneous nucleation of/on nanoparticles: a density functional study using the phase-field crystal model, (2007).

[30] M. Yang, L. Wang, W. Yan, Phase-field modeling of grain evolution in additive manufacturing with addition of reinforcing particles, Additive Manufacturing. 47 (2021) 102286. https://doi.org/10.1016/J.ADDMA.2021.102286.

[31] T. Pusztai, L. Rátkai, A. Szállás, L. Gránásy, Phase-Field Modeling of Solidification in Light-Metal Matrix Nanocomposites, Magnesium Technology. (2014) 455–459. https://doi.org/10.1002/9781118888179.CH83.

[32] S. Dewitt, S. Rudraraju, D. Montiel, W.B. Andrews, K. Thornton, PRISMS-PF: A general framework for phase-field modeling with a matrix-free finite element method, Npj Computational Materials. 6 (2020). https://doi.org/10.1038/s41524-020-0298-5.

[33] F. Han, B. Tang, H. Kou, J. Li, Y. Feng, Cellular automata simulations of grain growth in the presence of second-phase particles, Modelling and Simulation in Materials Science and Engineering. 23 (2015) 065010. https://doi.org/10.1088/0965-0393/23/6/065010.

[34] S.P. Gentry, K. Thornton, Simulating recrystallization in titanium using the phase field method, in: IOP Conference Series: Materials Science and Engineering, Institute of Physics Publishing, 2015. https://doi.org/10.1088/1757-899X/89/1/012024.

[35] M. Castro, Phase-field approach to heterogeneous nucleation, Physical Review B. 3 (2002) 67.

[36] T. Pusztai, G. Tegze, G.I. Tóth, Ĺ. Környei, G. Bansel, Z. Fan, Ĺ. Gŕńsy, Phase-field approach to polycrystalline solidification including heterogeneous and homogeneous nucleation, Journal of Physics Condensed Matter. 20 (2008). https://doi.org/10.1088/0953-8984/20/40/404205.

[37]  M. Yang, L. Wang, W. Yan, Phase-field modeling of grain evolutions in additive manufacturing from nucleation, growth, to coarsening, Npj Computational Materials. 7 (2021). https://doi.org/10.1038/s41524-021-00524-6.

[38]  E.E. Underwood, Stereology, or the quantitative evaluation of microstructures, Journal of Microscopy. 89 (1969) 161–180. https://doi.org/10.1111/J.1365-2818.1969.TB00663.X.

[39]  N. Moelans, B. Blanpain, P. Wollants, Quantitative analysis of grain boundary properties in a generalized phase field model for grain growth in anisotropic systems, Physical Review B - Condensed Matter and Materials Physics. 78 (2008). https://doi.org/10.1103/PhysRevB.78.024113.

[40]  S. Shahandeh, M. Militzer, Grain boundary curvature and grain growth kinetics with particle pinning, Philosophical Magazine. 93 (2013) 3231–3247. https://doi.org/10.1080/14786435.2013.805277.

[41]  N. Moelans, B. Blanpain, P. Wollants, Phase field simulations of grain growth in two-dimensional systems containing finely dispersed second-phase particles, (2005). https://doi.org/10.1016/j.actamat.2005.10.045.

[42]  N. Moelans, B. Blanpain, P. Wollants, An introduction to phase-field modeling of microstructure evolution, Calphad: Computer Coupling of Phase Diagrams and Thermochemistry. 32 (2008) 268–294. https://doi.org/10.1016/j.calphad.2007.11.003.

[43]  B. Sa, Comparison of Classical Nucleation Theory and Modern Theory of Phase Transition, J Adv Chem Eng. 7 (2017) 1. https://doi.org/10.4172/2090-4568.1000177.

[44]  D. Turnbull, Formation of Crystal Nuclei in LiquidMetals, Journal of Applied Physics. 21 (1950) 1022–1028. https://doi.org/10.1063/1.1699435.

[45]  J. Banerjee, S.H. Kim, C.G. Pantano, Elemental areal density calculation and oxygen speciation for flat glass surfaces using x-ray photoelectron spectroscopy, Journal of Non-Crystalline Solids. 450 (2016) 185–193. https://doi.org/10.1016/J.JNONCRYSOL.2016.07.029.

[46]  L. Ickes, A. Welti, C. Hoose, U. Lohmanna, Classical nucleation theory of homogeneous freezing of water: thermodynamic and kinetic parameters, Physical Chemistry Chemical Physics. 17 (2015) 5514–5537. https://doi.org/10.1039/C4CP04184D.

[47]  J.P. Simmons, C. Shen, Y. Wang, Phase field modeling of simultaneous nucleation and growth by explicitly incorporating nucleation events, Scripta Materialia. 43 (2000) 935–942. https://doi.org/10.1016/S1359-6462(00)00517-0.

[48]  W. Wu, D. Montiel, J.E. Guyer, P.W. Voorhees, J.A. Warren, D. Wheeler, L. Gránásy, T. Pusztai, O.G. Heinonen, Phase field benchmark problems for nucleation, Computational Materials Science. 193 (2021) 110371. https://doi.org/10.1016/j.commatsci.2021.110371.

[49]  J. Ding, D. Neffati, Q. Li, R. Su, J. Li, S. Xue, Z. Shang, Y. Zhang, H. Wang, Y. Kulkarni, X. Zhang, Thick grain boundary induced strengthening in nanocrystalline Ni alloy, Nanoscale. 11 (2019) 23449–23458.

[50]    BryanKinzer/zener-pinning-PRISMS-PF: Phase Field Zener Pinning Solidification PRISMS-PF Model, (2022). https://github.com/BryanKinzer/zener-pinning-PRISMS-PF/tree/main (accessed June 19, 2022).

[51]    C.J. Permann, M.R. Tonks, B. Fromm, D.R. Gaston, Order parameter re-mapping algorithm for 3D phase field model of grain growth using FEM, Computational Materials Science. 115 (2016) 18–25. https://doi.org/10.1016/J.COMMATSCI.2015.12.042.

[52]    Smith & Hashemi, Foundations of Materials Science and Engineering (4th ed.), (2006) 242.

[53]    Stainless Steel Powder 17-4PH Spherical - 3D Powder Hub, (n.d.). https://www.3dpowderhub.com/product/stainless-steel-powder-17-4ph/ (accessed June 28, 2022).

[54]    Yttrium Oxide Nanoparticles / Nanopowder (Y2O3, 99.99%, 30-45 nm), (n.d.). https://www.us-nano.com/inc/sdetail/748 (accessed June 28, 2022).

[55]    P. von Hippel, Skewness, International Encyclopedia of Statistical Science. (2011) 1340–1342. https://doi.org/10.1007/978-3-642-04898-2_525.

[56]    M. Apel, B. Böttger, J. Rudnizki, P. Schaffnit, I. Steinbach, Grain Growth Simulations Including Particle Pinning Using the Multiphase-field Concept, ISIJ International. 49 (2009) 1024–1029. https://doi.org/10.2355/isijinternational.49.1024.

[57]    D.J. Srolovitz, M.P. Anderson, G.S. Grest, P.S. Sahni, Computer simulation of grain growth-III. Influence of a particle dispersion, Acta Metallurgica. 32 (1984) 1429–1438. https://doi.org/10.1016/0001-6160(84)90089-0.

[58]    C.S. Smith, Grains, phases, and interphases: an interpretation of microstructure, Trans. Metall. Soc. AIME. 175 (1948) 15–51.

[59]    P.A. Manohar, M. Ferry, T. Chandra, Five Decades of the Zener Equation, ISIJ International. 38 (1998) 913–924.

[60]    I. Sinha, R.K. Mandal, Avrami exponent under transient and heterogeneous nucleation transformation conditions, Journal of Non-Crystalline Solids. 357 (2011) 919–925. https://doi.org/10.1016/j.jnoncrysol.2010.11.005.

[61]    R.D. Doherty, K. Li, K. Kashyup, A.D. Rollett, M.P. Andaraon, Computer Modelling of Particle Limited Grain Growth and Its Experimental Verification, 1989.

[62]    M.P. Anderson, G.S. Grest, R.D. Doherty, K. Li, D.J. Srolovitz, Inhibition of grain growth by second phase particles: Three dimensional Monte Carlo computer simulations, Scripta Metallurgica. 23 (1989) 753–758. https://doi.org/10.1016/0036-9748(89)90525-5.

[63]    K. Song, M. Aindow, Grain growth and particle pinning in a model Ni-based superalloy, Materials Science and Engineering: A. 479 (2008) 365–372. https://doi.org/10.1016/J.MSEA.2007.09.055.

[64]    B. Li, B. Qian, Y. Xu, Z. Liu, J. Zhang, F. Xuan, Additive manufacturing of ultrafine-grained austenitic stainless steel matrix composite via vanadium carbide reinforcement addition and

selective laser melting: Formation mechanism and strengthening effect, Materials Science and Engineering: A. 745 (2019) 495–508. https://doi.org/10.1016/J.MSEA.2019.01.008.

[65]   A. Jain, S.P. Ong, G. Hautier, W. Chen, W.D. Richards, S. Dacek, S. Cholia, D. Gunter, D. Skinner, G. Ceder, K.A. Persson, Commentary: The Materials Project: A materials genome approach to accelerating materials innovation, APL Materials. 1 (2013) 011002. https://doi.org/10.1063/1.4812323.

## Appendices

### Appendix I: Nomenclature

| | |
|---|---|
| **Nomenclature** | |
| $L_g$ | grain growth rate constant, m$^3$/(J·s) |
| $a_k$ | anisotropy coefficient |
| $l_g$ | grain boundary width, nm |
| $D_0$ | grain boundary pre-exponential constant, m$^4$/(J·s) |
| $Q_g$ | activation energy for grain boundary mobility, kJ/mol |
| $\Delta f_g$ | nucleation free energy barrier |
| $Q_D$ | activation energy for diffusion, kJ/mol |
| $m_g$ | interface constant, J/m$^3$ |
| $d$ | average grain size, μm |
| $H_L$ | latent heat, J/m$^3$ |
| $T$ | Temperature, K |
| $N_{sites}$ | areal density of nucleation sites, Nuclei/m$^2$ |
| $N_0$ | dimensionless nucleation sites |
| $x$ | pertaining to x-coordinate, nm |
| $y$ | pertaining to y-coordinate, nm |
| $t$ | time, ns |
| $N$ | number |
| $F$ | free energy of system, J |
| $f$ | elemental free energy, J/m$^3$ |
| $\vec{r}$ | spatial coordinate, nm |
| $J^*$ | areal nucleation rate, Nuclei/(m$^2$·s) |
| $\zeta_\theta$ | Contact angle function |
| $k_B$ | Boltzmann Constant, J/K |
| $h$ | Planck's Constant, J·m |
| $G^*$ | Critical Nuclei Free Energy, $J$ |
| $P$ | Nucleation Probability |
| $K_{HP}$ | Hall-Petch Constant, MPa/(μm)$^{0.5}$ |
| $G$ | Shear Modulus, GPa |
| $b$ | Burgers vector, nm |
| $L$ | interparticle spacing, nm |
| $M$ | Taylor Factor |
| $Cost$ | total alloy cost, $/kg |
| $P$ | price per kg, $/kg |
| $K$ | Zener equation proportional constant |
| $m$ | Zener equation exponential constant |
| $r_p$ | nanoparticle radius, μm |
| $s_{dz}$ | Grain Size Standard Deviation, % |
| *Greek Symbols* | |
| $\gamma$ | grain interaction coefficient |
| $\sigma_g$ | grain boundary energy, J/m$^2$ |
| $\sigma_p$ | liquid melt/solid interfacial energy, J/m$^2$ |
| $\sigma_{HP}$ | Hall-Petch Strength, MPa |
| $\sigma_{Or}$ | Orowan Strength, MPa |
| $\sigma_y$ | yield strength, MPa |
| $\sigma_0$ | base matrix strength, MPa |
| $\theta$ | nuclei contact angle |
| $\eta$ | order parameter |
| $\kappa_g$ | gradient coefficient, J/m |
| $\varphi$ | nanoparticle order parameter |
| $v$ | Poisson's ratio |
| $\phi$ | area fraction |
| $\Theta$ | mechano-economic metric, MPa/($/kg) |
| $\Gamma_z$ | fraction of nanoparticles on grain boundaries |
| *Other Subscripts* | |
| $i$ | general index |
| $j$ | general index |
| $z$ | predicted grain size, Zener |
| *np* | pertaining to nanoparticles |
| SS316L | stainless steel 316L |
| *nuc* | pertaining to nucleation |
| 0 | pertaining to element |
| *bulk* | pertaining to bulk |
| *grad* | pertaining to gradient |
| m | melting point |

## Appendix II: Explanation of PRISMS-PF Phase-field Grain Remapping

During the simulation if two grains of the same order parameter come close to touching, the remapping algorithm switches one of the order parameters to avoid artificial coalescence to maintain a realistic microstructure. With grain remapping two grains may collide and merge before they can be reassigned, so simulations are considered acceptable if less than 1% of grains artificially merged. The highest number of grains generated for a simulation is over 400, but $N$ = 16 grain orientations is found sufficient to simulate realistic microstructures, which reduces computation time and memory use [60]. This is less than half the required orientations of other similar nanoparticle grain growth simulations [37,49].

Appendix III: Explanation of MATLAB Grain Analysis Program

For grain size analysis, a custom MATLAB grain image analysis tool is used to count the number of distinct grains, individual grain areas, and fraction of nanoparticles on grain boundaries. Since the locations of each nanoparticle are known, a subdomain, four times the individual particle radius, containing the nanoparticle is inspected. Since each grain at a grain boundary has a distinct color, the number of colors present in the image is checked. If two or more colors are present (filtering out the color of the nanoparticle) that indicates the presence of a grain boundary and the particle counts towards the pinning fraction.